\documentclass[sigconf]{acmart}

\AtBeginDocument{%
  }

\setcopyright{acmlicensed}
\copyrightyear{2025}
\acmYear{2025}
\setcopyright{acmlicensed}\acmConference[CHI '25]{CHI Conference on Human Factors in Computing Systems}{April 26-May 1, 2025}{Yokohama, Japan}
\acmBooktitle{CHI Conference on Human Factors in Computing Systems (CHI '25), April 26-May 1, 2025, Yokohama, Japan}
\acmDOI{10.1145/3706598.3713248}
\acmISBN{979-8-4007-1394-1/25/04}

\begin{document}

\title{A Matter of Perspective(s): Contrasting Human and LLM Argumentation in Subjective Decision-Making on Subtle Sexism}

\author{Paula Akemi Aoyagui}
\email{paula.aoyagui@mail.utoronto.ca}
\orcid{0000-0001-7779-2403}
\affiliation{%
  \institution{University of Toronto}
  \city{Toronto}
  \country{Canada}
}

\author{Kelsey Stemmler}
\email{kelsey@stemmler.tech}
\affiliation{%
  \institution{University of Toronto}
  \city{Toronto}
  \country{Canada}
}

\author{Sharon Ferguson}
\email{sharon.ferguson@mail.utoronto.ca}
\affiliation{%
  \institution{University of Toronto}
  \city{Toronto}
  \country{Canada}
}

\author{Young-Ho Kim}
\email{ygho.kim@navercorp.com}
\affiliation{%
  \institution{NAVER AI Lab}
  \country{Republic of Korea}}

\author{Anastasia Kuzminykh}
\email{anastasia.kuzminykh@utoronto.ca}
\affiliation{%
 \institution{University of Toronto}
 \city{Toronto}
 \country{Canada}}

\renewcommand{\shortauthors}{Aoyagui et al.}

\begin{abstract}
 In subjective decision-making, where decisions are based on contextual interpretation, Large Language Models (LLMs) can be integrated to present users with additional rationales to consider. The diversity of these rationales is mediated by the ability to consider the perspectives of different social actors; however, it remains unclear whether and how models differ in the distribution of perspectives they provide. We compare the perspectives taken by humans and different LLMs when assessing subtle sexism scenarios. We show that these perspectives can be classified within a finite set (perpetrator, victim, decision-maker), consistently present in argumentations produced by humans and LLMs, but in different distributions and combinations, demonstrating differences and similarities with human responses, and between models. We argue for the need to systematically evaluate LLMs’ perspective-taking to identify the most suitable models for a given decision-making task. We discuss the implications for model evaluation.
\end{abstract}




\maketitle

\textit{Warning:} This paper contains examples and quotes that may be upsetting or offensive, specifically about gender bias.

\section{Introduction}
Subjectivity in AI-assisted decision-making presents unique challenges and opportunities for human-AI collaboration. The ultimate aim of human-AI collaboration is to leverage the complementary strengths of both AI and humans \cite{bansal2021does, hemmer2021human}; however, conventional approaches to machine learning frame decision-making predominantly as an objective task that can be resolved to a unilateral ground truth judgment, often overlooking subjective decision-making. In subjective tasks, the decision is based on contextual interpretation and multiple decision outcomes can be applicable \cite{aoyagui2024exploring}. Correspondingly, AI systems, in this case, can act as complementary decision-making tools that offer diverging \cite{lai2021towards, ferguson2024just, ferguson2023something} or critical perspectives (e.g., acting as devil's advocate \cite{chiang2024devil} or provocateur \cite{Sarkarchallenge}), which is commonly implemented by using Large Language Models (LLMs) \cite{Conflict.resolution, gao2024collabcoder, vowels2024,dhillon.creative}.

Although human-AI collaboration in subjective decision-making finds a wealth of applications across various domains (e.g., professional and interpersonal relationship advice \cite{vowels2024}, creative writing \cite{dhillon.creative}, or conflict resolution \cite{Conflict.resolution}), recent explorations focused on online content moderation \cite{frenda2024perspectivist,ferguson2023something,kumarage2024harnessing}, where AI, in general, \cite{lai2022human}, and LLMs specifically \cite{kumar2023watch, zhang2023efficient, schick2021self, mishra2023exploring}, assist humans in identifying hate speech targeting specific groups or populations \cite{albadi2018they, alfina2017hate, fortuna-etal-2019-hierarchically, demus2022comprehensive}. While some toxic content is relatively straightforward to detect (e.g., when explicit derogatory language is used \cite{gehman2020realtoxicityprompts, cao2020deephate}), more implicit and context-dependent hate speech with social and cultural nuances \cite{garg2023handling, yin2021towards, kim2020intersectional} 
poses significant challenges for automatic detection \cite{wang2022toxicity, oliva2021fighting, diaz-etal-2022-accounting} and LLM-based classification methods \cite{kumar2023watch}. 
Subtle sexism (i.e., discrimination based on gender) exemplifies such nuanced forms of discrimination, as its interpretation can vary across contexts \cite{Benokraitis1997,hammond2018benevolent, swim&cohen1997, hall2016they, barreto2005burden} and depends heavily on the decision-maker's values and experiences \cite{mitamura2017value}. Notably, AI-provided rationales and arguments have proven particularly valuable in the collaborative assessment of subtle sexism scenarios \cite{ferguson2023something}.

While AI recommendations in objective tasks can be evaluated with metrics for assessing ``objective'' performance  (e.g., accuracy, robustness
\cite{liang2023HELM,fok2023search}), the nature of subjective decision-making necessitates a more nuanced approach for navigating the validity of multiple possible interpretations \cite{baron2014groundless}. 
An important caveat of such arguments, rationales, and explanations is that they are, by definition, \textit{biased}. This inherent bias underscores the necessity of considering diverse perspectives within subjective decision-making scenarios. In this context, AI's collaborative input serves as an additional, potentially biased perspective for consideration \cite{ferguson2023something}. The ability to understand a situation from another person's viewpoint, known as \textit{perspective-taking} \cite{batson1997perspective, healey2018cognitive}, has been shown to enhance an LLM's reasoning capabilities when used as a prompting mechanism \cite{wilf2023think}. However, requiring users to prompt specific perspectives would undermine the collaborative purpose of using AI to surface perspectives that the user hadn't considered. Therefore, a model should be capable of autonomously providing complementary, albeit potentially biased, perspectives. This raises important questions about how models differ in the diversity and distribution of perspectives they offer, and whether these perspectives truly complement (rather than redundantly mirror) human reasoning. Moreover, it highlights the need to critically evaluate the biases present in both AI-generated and human-generated arguments, ensuring a balanced and comprehensive decision-making process.

This paper investigates the alignment of perspectives between humans and LLMs in a subjective decision-making task focused on subtle sexism. Our study employs a diverse set of LLMs, including an open-source model (Meta's Llama 3.1, released in 2024) and multiple iterations of OpenAI's GPT series \cite{openai2023gpt4} (GPT-3, GPT-3.5, and GPT-4). This selection allows for comparisons between open- and closed-source models, as well as between different versions within the GPT series. We prompt both human participants and these LLMs to assess a collection of scenarios sourced from the internet, specifically designed to explore subtle forms of sexism. Crucially, our methodology does not treat the human dataset as ground truth, acknowledging the inherently subjective nature of the scenarios under evaluation. Instead, our comparative analysis of human and LLM datasets aims to identify the diversity and distribution of perspectives chosen, indicating potential for complementary performance, where the models could offer novel viewpoints for human decision-makers to consider. This approach provides insights into the potential of LLMs to enhance and diversify human reasoning in complex, subjective decision-making contexts.

Our study revealed that both humans and LLMs employ similar assessment parameters: Stances (\emph{sexist}, \emph{not sexist}, \emph{depends}, or \emph{no stance}) and Perspectives (\emph{victim}, \emph{perpetrator}, and \emph{decision-maker}). At the same time, while the Stances were mutually exclusive (i.e., an assessment could only be categorized with one option), the Perspectives could be combined (i.e., the same response could have both the \emph{victim}'s and the \emph{perpetrator}'s). However, our mixed-methods analysis uncovered distinct tendencies in how these parameters are used across different datasets. For instance, GPT-3 outputs were less likely to adopt a \emph{sexist} stance, while GPT-3.5 and GPT-4 responses often demonstrated counterfactual reasoning through increased use of the \emph{depends} stance. GPT-4 demonstrated a trend towards impartiality, while the open-source Llama 3.1 more frequently argued using the \emph{victim}'s perspective and offered unsolicited advice to make the scenario `less offensive'. Notably, we observed evolving approaches to the \emph{decision-maker}'s perspective, from anthropomorphic characteristics in earlier models to the explicit acknowledgment of AI limitations in newer versions. When comparing human and LLM datasets, we found GPT-3 had the most similarities to human responses, notably, in low consistency in assessments (i.e., the responses to a scenario had more variability), in fewer Perspective combinations (i.e., more simplistic answers) and more use of personal experiences (including GPT-3 outputs mimicking human experiences to the point of self-identifying as a woman). Overall, our results point to a gradual refinement in justification present in the model responses, as newer model iterations provided more comprehensive and nuanced analyses contemplating multiple perspectives of the different actors in the scenario.

These findings highlight the importance of perspective diversity in model selection for subjective decision-making tasks. Our work demonstrates that models within the same family (GPT-3, GPT-3.5, GPT-4) and across different types (Llama 3.1 vs. GPTs) exhibit unique tendencies in adopting Stances and Perspectives. This variability is crucial for designing effective LLM-assisted solutions in subjective decision-making, as the model's utility in human-AI collaboration depends on its ability to offer complementary perspectives. We contextualize these findings within recent discussions in the Human-Computer Interaction (HCI) community regarding the diverse roles of AI systems in decision-making \cite{ma2024beyond} to not only provide the most accurate recommendation but, in some use cases, to propose adversarial viewpoints \cite{chiang2024devil} and challenge the human counterpart \cite{Sarkarchallenge, sarkar2024large} to foster critical thinking \cite{danry2023don}. This shift from a traditional recommender-system role to `devil's advocate' or `challenger' roles will require new parameters to develop and evaluate models, especially when applied to non ground truth and value-based scenarios. Thus, we propose Stances and Perspectives as criteria to be leveraged for fine-tuning and/or for measuring model performance in the aforementioned use cases. Furthermore, our gender-inclusive categorizations of Perspectives contribute to the ongoing efforts to evaluate LLMs, potentially enhancing existing benchmark techniques for assessing a model's perspective-taking abilities in ambiguous social interactions. Therefore, the contributions of this work are twofold: 1) identifying the similarities and differences in Stance and Perspectives in responses from humans, GPT-3, GPT-3.5, GPT-4 and Llama 3.1 when assessing subtle sexism scenarios, as an example of subjective decision-making, and 2) proposing \emph{Stance} and \emph{Perspectives} as criteria to inform the design and evaluation of LLMs when applied to human-AI collaborative decision-making settings where there is no ground truth.

\section{Related Work}

\subsection{Human-AI Subjective Decision-Making}

Recent advances in AI technologies have prompted an active development of the field of human-AI collaboration and the exploration of its effectiveness \cite{dakuo.et.al,arous.et.al, cai2019hello}. For example, \citet{Jacobsen.et.al} demonstrated how human-AI collaboration can increase both the actual and the perceived effectiveness of one's performance when completing a task.
Aiming to leverage the best of humans' and AI's skills \cite{steyvers2023three,jarrahi2018artificial}, this collaboration can take many forms, e.g., humans offering feedback to train the model \cite{dellermann2021future} or humans and AI working as a team towards a common goal \cite{zhang2021ideal, kelly2023capturing}. However, the most common format for human-AI collaboration revolves around an AI system providing recommendations, while the human counterpart remains the ultimate decision-maker \cite{guerdan2023ground, zhang2020effect, fogliato2022goes}. 

Conventional approaches to human-AI collaborative decision-making have predominantly focused on decision tasks that have one ultimately correct outcome \cite{lai2021towards}, i.e., ground truth, (e.g., predictions for medical diagnosis \cite{cai2019hello} or financial assessment \cite{binns2018s}). 
The collaborative effectiveness of objective decision-making is often viewed through the lens of optimal accuracy in complementary performance, i.e., where human-AI joint accuracy is higher than either the human or the AI working on its own \cite{bansal2021does, hemmer2021human, donahuecomplementarity}. 
However, not all decision-making is objective, and often decisions have to be made about nuanced and contextually-dependant scenarios.  
Correspondingly, the possibility of AI collaborative assistance in application to subjective decision-making, where an AI system might aid by providing additional points of view \cite{hemmer2021human}, has been recently gaining more attention \cite{Schaekerman.et.al,Inkpen.et.al} in a variety of domains, e.g., legal decisions \cite{hayashi.et.al}, music production \cite{nicholls2018collaborative}, or recognition of social prejudice \cite{ferguson2023something}. \citet{vowels2024}, for example, showed the effectiveness of chatbots in providing relationship-related advice, and \citet{dhillon.creative} explored the effects of scaffolding from LLMs on the creative writing process. Other examples include LLMs offering alternative strategies to manage conflict resolution \cite{Conflict.resolution} or facilitating agreement between human annotators in qualitative coding for academic research \cite{gao2024collabcoder}.

\subsection{Evaluating LLMs in Subjective Decision-Making}

Past literature has established that effective assistance in subjective decision-making requires a system to consider divergent understandings and interpretations \cite{frenda2024perspectivist, sorensen2024roadmap}, both in collaborating with an individual human \cite{ferguson2023something}, and in AI-assisted group decision-making where the agent must consider different perspectives and opinions of each individual in the group (e.g., multiple family members making vacation plans \cite{delic2024supporting}). To describe the ability to consider different perspectives, research in LLMs has borrowed \cite{xu2024walking,wilf2023think} the perspective-taking concept \cite{batson1997perspective} from cognitive psychology, which refers to the ability to interpret another person's point of view as separate from one's own \cite{healey2018cognitive, batson1997perspective}. Note that perspective-taking happens from a third-person perspective (i.e., imagining what another person is thinking) and is conceptually different from role-playing \cite{lu2024role-play} or personas \cite{ha2024clochat}, which assumes a first-person perspective. 
For example, \citet{xu2024walking} successfully used perspective-taking as a prompting strategy to remove toxicity and biases in LLM responses. \citet{wilf2023think} demonstrated that perspective-taking is a relevant skill to improve social reasoning in LLMs. Social reasoning is a concept related to the Theory of Mind (ToM) \cite{ullman2023largetom, y2022largetom}, which encompasses the ability to understand another person's intentions, desires, and beliefs to infer how these factors influence the person's behaviour. Past work evaluating ToM capabilities in LLMs often compares how models and humans perform with mixed results, depending on the task observed. For example, \citet{gandhi2024understanding}'s proposed benchmark compared five commercial and open-sourced LLMs and found that only GPT-4 performed closely to humans, albeit with some limitations. On the other hand, \citet{wang2024evaluating} found that GPT-4 underperformed in social intelligence tasks compared to humans.

Indeed, benchmarking is a popular approach for assessing LLM capabilities, allowing researchers to compare how different models and versions perform, either in a specific task or more holistically \cite{liang2023HELM}. However, given that the purpose of AI in subjective decision-making is to bring up perspectives that the user had not considered \cite{lai2021towards, ferguson2024just}, solely assessing the model's performance against human benchmark seems counterproductive, since the goal of this collaboration is complementarity rather than similarity of reasoning. Thus, the effectiveness of LLMs for assisting in subjective decision-making depends on whether the diversity of perspectives they offer truly complements human reasoning \cite{ferguson2024just, ferguson2023something}.
And yet, we currently lack an understanding of how to evaluate the distribution and frequency of the perspectives LLMs tend to display in their outputs \cite{frenda2024perspectivist} and whether these distributions differ between models. 

\section{Methods}

\subsection{Data Collection}
This study leverages six datasets: one scenario, one human, and four LLM-generated (GPT-3, GPT-3.5, GPT-4 and LLama 3.1). We describe the data collection process for each of them below. 

\subsubsection{Scenarios Dataset} Existing datasets for detecting toxicity and hate speech \cite{waseem-hovy-2016-hateful, mathew2020hatexplain, mollas2020ethos, gehman2020realtoxicityprompts} contain examples of overt sexism that were not adequate for this study since our focus is on ambiguous scenarios. Hence, we chose to collect our dataset of subtle sexism scenarios through online discussion forums\footnote[1]{\url{www.reddit.com}, \url{www.everydaysexism.com}, \url{www.twitter.com}}. 
The search was conducted in June-July 2021 using pre-determined keywords such as \textit{``subtle sexism'', ``everyday sexism'', ``why is this sexist?'', ``is this sexist?'', and ``experiences of sexism''}. As a result, we have 101 scenarios of subtle sexism that are open to interpretation.

\subsubsection{GPT-3, GPT-3.5 and GPT-4 Datasets} 
To compare responses from different model versions of the GPT series, we chose GPT-3 \cite{GPT-3documentation} (dataset collected in 2021), GPT-3.5 and GPT-4 \cite{GPT-4techreport} (datasets collected in 2023). The same protocol was followed for consistency, with parameters set for default, except for (1) the maximum number of tokens and (2) temperature (amount of randomness in model output). The former was set to 240 words to gather fully-formed answers without unrelated text (especially needed for GPT-3). For temperature, up until GPT-3.5, the scale ranged from 0 to 1 (with default at 0.7). Then, on GPT-4 the scale changed to 0 to 2 (default at 1) \cite{temperatureOpenAI, communityOpenAI2023}. Hence, we used the default temperature of 0.7 for GPT-3 and GPT-3.5, and the default of 1 for GPT-4. For prompting, we used the \textit{Scenarios Dataset} in question-answer format (see examples in Appendix \ref{appendixquestion-answer}). Since the quality of outputs from these models can vary, particularly in GPT-3, each scenario was then used to prompt all models three times, obtaining three different responses.

\subsubsection{Llama 3.1 Dataset}
To compare responses across different model types, we chose to collect a dataset using an open-source model, Llama 3.1. The data was collected in 2024, using the latest available Llama model at that time (Llama 3.1-8b \cite{llama}), using the same protocol described above, for consistency. We set the maximum number of tokens initially to 240, same as for all GPT model versions. However, we found that all of the Llama 3.1 responses came up truncated and, therefore, incomplete. We then set the maximum number of tokens to 500 and successfully retrieved all fully-formed answers. The temperature value range for Llama 3.1 is 0 to 2, with default at 0.6, which is what we used to follow a similar protocol described above. For prompting, we used the Scenarios Dataset in question-answer format and prompted the model three times to obtain three different responses, to be consistent with the data we collected previously for the GPT models.

\subsubsection{Human Dataset}
Our sixth and final dataset was collected using the online survey platform, SurveyMonkey. Participants were recruited via the Target Response function\footnote[4]{\url{https://www.surveymonkey.com/market-research/solutions/audience-panel/}} 
and we chose to balance demographic parameters based on the United States Census, where responses were collected. There was also intentional balance for gender (48.87\% male, 55.13\% female), age (13\% 18-29, 41\% 30-44, 14\% 45-60 and 32\% >60 years old) and household income (\$0-\$200,000). The study was approved by the institution's Research Ethics Board and no personally identifiable information was collected from participants. Additionally, all 104 participants provided informed consent and received compensation. Each participant was shown a random selection of 5-6 scenarios (of the 101 from the \textit{Scenarios Dataset}) and was asked: ``Is this scenario sexist, why or why not?''. No participant saw the same scenario twice.

\subsection{Data Analysis}

\subsubsection{Inclusion Criteria} We started by inspecting the datasets collected from humans and GPTs (GPT-3, GPT-3.5 and GPT-4) to remove nonsensical responses (i.e., solely gibberish or unrelated to the question) or incomplete (i.e., did not contain an assessment and a rationale to support it). It is important to note that the quality of the rationale was not under scrutiny; it only had to be present. Two researchers worked collaboratively in this inclusion analysis. From 303 collected responses from GPT-3, 74 were removed, resulting in 229 valid responses (mean = 2.26 valid responses per scenario; mode = 2 valid responses per scenario; standard deviation = 0.74). Similarly, the human dataset had 209 responses removed (out of 516 collected), resulting in 307 valid responses for analysis for the same reasons described above (mean = 3.05 valid responses per scenario; mode = 2 responses per scenario; standard deviation = 1.89). See examples of excluded examples in Appendix \ref{app-invalid}. Notably, no responses were disqualified in the \textit{GPT-3.5 Dataset}, the \textit{GPT-4 Dataset} or the \textit{Llama 3.1 Dataset}.

In the next subsections, we discuss Researcher Positionality, describe the iterative coding process, and provide detailed definitions of the coding schemes.

\subsubsection{Researcher Positionality Statement}
In line with the practice of reflexivity \cite{corlett2018reflexivity}, we discuss how our identities and perspectives inform this work, particularly the qualitative analysis. The researchers directly involved in this process come from different socio-cultural backgrounds and identify as women who have worked in traditionally male-dominated fields in Europe, North America and South America, meaning our own experiences of sexism may inform our analysis. Nonetheless, past work argues that gender does not necessarily determine a person's stance on sexism, but their personal values, beliefs, experiences and diverse socio-cultural background also play a role \cite{mitamura2017value}.

\subsubsection{Qualitative Coding} Next, we were interested in two elements of an assessment of subtle sexism --- the Stance (i.e., the answer to: ``is this scenario sexist?") and the Perspectives (i.e., the point of view of different actors in the scenario used for argumentation). We chose qualitative coding \cite{saldana2021coding, adams2008qualititative, williams2019art} since past work shows automated methods might miss nuanced elements of sexism in text \cite{sheng2019babysitter, kocielnik2023biastestgpt, kotek2023gender}. We followed a systematic and iterative process \cite{glaser2017discovery, lazar2017research} to identify coding schemes for Stance and Perspective, to be described in the next sections.

\subsubsection{Coding Scheme: Stance} We used a top-down approach, and after careful discussion of the definitions and boundaries for each code following the process described below, we came to the definitions in Table \ref{tab:CodingScheme_Stance}. Two researchers first used one-quarter of each dataset for training (human, GPT-3, GPT-3.5, GPT-4), and then coded one-quarter separately (again, from each dataset) to measure agreement. The training was split into multiple iterations to ensure no new \emph{Stances} emerged. Raw agreement ranged from 89\% of examples at the low end (human dataset), to 96\% of the high end (GPT-3.5 dataset), representing sufficient agreement. The remaining half of each dataset was split equally among the two coders. The Llama 3.1 dataset was collected and coded later, following the same protocol for training and agreement---three researchers (two of them outside coders, who did not participate in the creation of the coding scheme) used the same coding scheme, no new \emph{Stances} were identified and agreement was 93\% before coders worked separately.

For Stance, responses from all datasets (human and LLMs) were classified using one of the four codes: \textit{sexist}, \textit{not sexist}, \textit{depends} or \textit{no stance} (examples in Table \ref{tab:CodingScheme_Stance}). Both the \textit{sexist} and \textit{not sexist} codes were easily categorized, as the Stance was often explicitly stated. For example: \textit{``The statement you mentioned can be considered \textbf{sexist} because it perpetuates a stereotype about women taking a long time to shop or buying extra stuff."} (GPT-3.5). And: \textit{``\textbf{not sexist}. True. Most women like to accumulate stuff"} (Human). The \textit{depends} code was used whenever a factor or condition influenced or determined the assessment, most often the intentions behind the action. In most cases, the word `depends' was included in the answer, or the counterfactual reasoning was evidenced by the structure `if this, then that', as in \textit{``It's only sexist \textbf{if the person} that said it was being sexist (which they may not have been) or \textbf{if the person} saying it is sexist. \textbf{If it was just a joke} between friends about something that happened, that's different than if someone says that women who do too much shopping are annoying"} (GPT-3). In this example, dependency is predicated upon the intentions of the person being accused (i.e., if they meant to be sexist or meant it as a joke). Additionally, nuances in context were also highlighted as the factor that would tip the scales: \textit{``However, the \textbf{context and intention} behind the actions could determine whether it is sexist or not.”} (GPT-4). Lastly, the \textit{no stance} code differs from \textit{depends} because no definitive judgement in stance is reached. At times, the response also stated further details would be required to make a final call: \textit{``Without further context or information, it is \textbf{difficult to definitively determine} if the statement is sexist or not"} (GPT-3.5).

\renewcommand{\arraystretch}{1.5}
\begin{table*}[h]
\caption{Coding scheme for Stance present in each assessment}
\resizebox{\textwidth}{!}{%
\label{tab:CodingScheme_Stance}
\begin{tabular}{p{1.5cm}p{4cm}p{10cm}}
 Stance    & Definition                                                                                                             & Example                                                                                                                                                                                                                                                                                                \\ \hline \emph{Sexist}         & Yes, the scenario is \newline considered sexist                                     & \textit{``This is sexist. No matter the familiarity of the parties, such a statement unfairly characterizes and diminishes women.'' (Human)}                                                                                                             \\ \cline{1-3} 
\emph{Not Sexist}     & No, the scenario is \newline considered not sexist                                   & \textit{``To me it's not. This is the type of language that people use to mean something positive." (GPT-3)}                                                                                                                                             \\ \cline{1-3} 
 \emph{Depends}        & One or more conditions would change the final assessment            & \textit{``If being chivalrous is done out of a genuine desire to show respect and kindness, it can be a positive trait. However, if it is done to reinforce patriarchal attitudes or stereotypes, it can be seen as sexist.'' (Llama 3.1)}                                         \\ \cline{1-3} 
\emph{No Stance}      & The response purposefully does not include a final assessment         & \textit{``It's possible the assistant addressed you because you were standing and therefore seemed more approachable. However, without more context, it's hard to definitively label this situation as sexist.'' (GPT-4)}                             \\ 
\hline
\end{tabular}
}
\end{table*}

\subsubsection{Coding Scheme: Perspectives}
We used bottom-up coding to identify the Perspectives (i.e., the point of view from which actors in the scenario were used to justify a rationale)---starting with open coding the human dataset of responses followed by a thematic analysis methodology \cite{braun2006using}. Two authors began by identifying open codes in one-fourth of the human dataset together, after which codes were discussed with another author and consolidated or rejected if they did not present a distinct perspective. The iterative training process required adding to the coding scheme if a new perspective emerged, applying this to past instances. Following this, two researchers independently coded another quarter of the human dataset, resulting in a 78\% raw agreement rate, and the remainder of the dataset was coded independently. Following the Perspective coding on the human dataset, the next step was to apply this coding scheme to the LLM Datasets (GPT-3, GPT-3.5, GPT-4 and Llama 3.1), to identify whether the same Perspectives were present, or if these models produced any new Perspectives not present in the human dataset. We did not identify any new perspectives in the model datasets. The same process was followed: one-quarter of each dataset was used to train, one-quarter measured agreement, and half was split between two coders independently. The resulting raw agreement for the GPT-3, Llama 3.1, GPT-4 and GPT-3.5  were 87\%, 89\%, 94\% and 98\% respectively.

\renewcommand{\arraystretch}{1.5}
\begin{table*}[h]
\caption{Coding scheme for Perspectives present in each assessment}
\label{tab:CodingScheme2}
\resizebox{\textwidth}{!}{%
\begin{tabular}{p{1.5cm}p{4cm}p{10cm}}
Perspective   & Definition                                                                                                             & Example                                                                                                                                                                                                                                                                                                \\ \hline
 \emph{Victim}         & Negative outcome to a person(s) or group                                & \textit{``Yes this is they took advantage of her work ethic and made a joke out of it to make her feel inferior'' (Human)}                                                                                                                               \\ \cline{1-3} 
 \emph{Perpetrator}    & Intention of the person(s) or group being accused of subtle sexism & \textit{``The statement itself is not inherently sexist, as it merely describes a man offering to perform certain actions for women. However, the context and intention behind the actions could determine whether it is sexist or not.''(GPT-4)} \\ \cline{1-3} 
\emph{Decision-maker} & Author's (i.e. GPT-3, GPT-3.5, GPT-4, humans) beliefs or \newline experiences & \textit{`` I am not sure I would call it "sexist", but it is kind of a weird message (about being emotional) to send to a female candidate.''} (GPT-3)                                                                                                   \\ \hline
\end{tabular}
}
\end{table*}

Our coding scheme revealed a consistent set of three Perspectives present in all datasets (human and LLMs). Definitions and examples are outlined in Table \ref{tab:CodingScheme2}.

The \emph{victim's} perspective was identified any time a negative outcome to the victim (the person to whom the potentially sexist act/statement was done/said) was mentioned in the response. It is important to note that the \emph{victim's} perspective is \emph{not} `pro-victim' in the scenario; instead, the coding is more nuanced, based on literature \cite{mitamura2017value} that points to the importance of considering victim outcomes when assessing sexism. An example of negative victim outcome that was present in all datasets refers to perpetuating harmful biases, as seen here: \textit{``The language used in the statement perpetuates negative stereotypes about women and implies entitlement and disrespect towards them."} (GPT-3). However, the \emph{victim's} perspective can also be used to conclude that the scenario is not sexist (i.e., there was no harm caused to the victim because of their gender). As exemplified in \textit{``The language used is not derogatory towards women, and the author appears to be critiquing the behaviour of the harasser rather than making a statement about women in general."} (Llama 3.1).
 
The same distinction applies to the \emph{perpetrator's} perspective, as it is \emph{not} `pro' nor against the perpetrator. This code is used when the author considers how the intentions, behaviours or socio-cultural context of the person(s) accused of sexism in the scenario played a role in the assessment, as seen on \textit{``This can be interpreted as sexist, though it likely comes from a place of chivalry and politeness, not intending harm."} (Llama 3.1).

The \emph{decision-maker's} perspective is present when the author overtly states their opinions, beliefs or experiences. While all responses can be implicitly interpreted as the author's opinions, this code is used in explicit instances where they highlight themselves as an agent in the decision. In particular, GPT-3 refers to personal experiences as part of justification, at times creating an uncanny valley effect \cite{mori2012uncanny}: \textit{``...I am a mother with three daughters...I am a feminist in the correct sense of the word. I believe in gender equality."} (GPT-3). Another characteristic of the \emph{decision-maker} perspective present in all datasets was the use of personal pronouns---but for different purposes. While humans, GPT-3 and Llama use it to explicitly mark their personal opinions or beliefs, GPT-3.5 and GPT-4 do the opposite---using ``I", ``me" or ``mine" to make a disclaimer about their inability to have personal opinions, beliefs or feelings:  \textit{``As an AI, I don't have feelings, but I can tell you that the phrase `men at work' can be interpreted in different ways.."} (GPT-4)

Further, we note that our \emph{Perspective} coding is gender-inclusive --- as \emph{perpetrators} and \emph{victims} can be of any gender. In fact, in some instances, both men and women were found to be the \emph{victims} of sexism: \textit{``This is sexist both because it assumes \textbf{men automatically have no sexual self-control} and that \textbf{women are obligated to act a certain way} to avoid danger"} (Human). While we did not find any responses referencing gender-non-normative examples in any dataset collected (e.g., non-binary or transgender), the coding scheme would be applicable.

\subsubsection{Quantitative and Qualitative Analyses:} Lastly, we adopted a mixed-methods \cite{TurnhoutMixedMethods} approach, combining qualitative and quantitative analyses to identify similarities and differences in the responses from humans and LLMs. For quantitative analysis, we began by tabulating the frequency of each \textit{Stance} and \textit{Perspective}. To determine whether the frequencies significantly differed, we conducted a one-way ANOVA analysis, with author type (human, GPT-3, GPT-3.5, GPT-4 and Llama 3.1) as the independent variable, and \emph{Perspective} or \emph{Stance} as the dependent variable. After filtering each dataset to only coherent responses, each dataset had a different number of total outputs (see Table \ref{tab:stance-total}). To compare whether each dataset displayed different Stances and Perspectives \textit{in response to the same scenario}, we averaged the Stances and Perspectives within each dataset at the scenario level. The resulting data then represented the percent chance that the model (or human) would show a specific stance if shown a specific scenario. For example, if GPT-3 outputs for scenario 1 displayed the Stances \emph{sexist}, \emph{sexist}, and \emph{not-sexist}, the resulting averages would show that there is a 66\% chance that GPT-3 displays the \emph{sexist} stance if prompted with scenario 1, and a 33\% chance the answer would be \emph{not sexist}. Averaging Stances and Perspectives at the scenario level also allows for an equal sample size across all datasets, which means the assumptions behind the ANOVA test are more likely to hold \cite{field2012discovering}.
Next, we applied the Bonferroni post-hoc analysis to identify which authors differed. 

For qualitative data analysis, we used content and thematic analyses \cite{braun2006using, adams2008qualititative} to identify differences and similarities in Stance and Perspectives within and across the human and LLM responses. We decided on an inductive approach, starting from the raw data (responses from all datasets that were assigned a certain Stance or Perspective) and annotating to identify codes, and then themes. The process was iterative, with repeated comparisons to cross-check and further analyze findings. We were particularly interested in understanding how human participants used Perspectives to justify their choice in Stance and then contrasted them to LLM outputs. This allowed for a qualitative exploration into the articulation and presentation styles, thus enhancing the quantitative data analysis described above. For example, by examining each response where the \emph{decision-maker} Perspective was present across datasets, we were able to identify clear differences in the use of personal pronouns, as we report in the following \textit{Results} section.

\section{Results}
In this section, we compare the argumentative Stance in responses across datasets to investigate the tendencies displayed by humans, GPT-3, GPT-3.5, GPT-4, and Llama 3.1. We then compare how Perspectives were used across the explored datasets.
Since the assessed scenarios are open to interpretation, the diversity in responses is expected within and across datasets. Therefore, we did not consider the human dataset as a benchmark that LLMs should aim to match; instead, we treated it as an equal dataset for comparison.

\subsection{Stance}
\begin{figure*}
    \centering
    \includegraphics[width=0.6\linewidth]
    {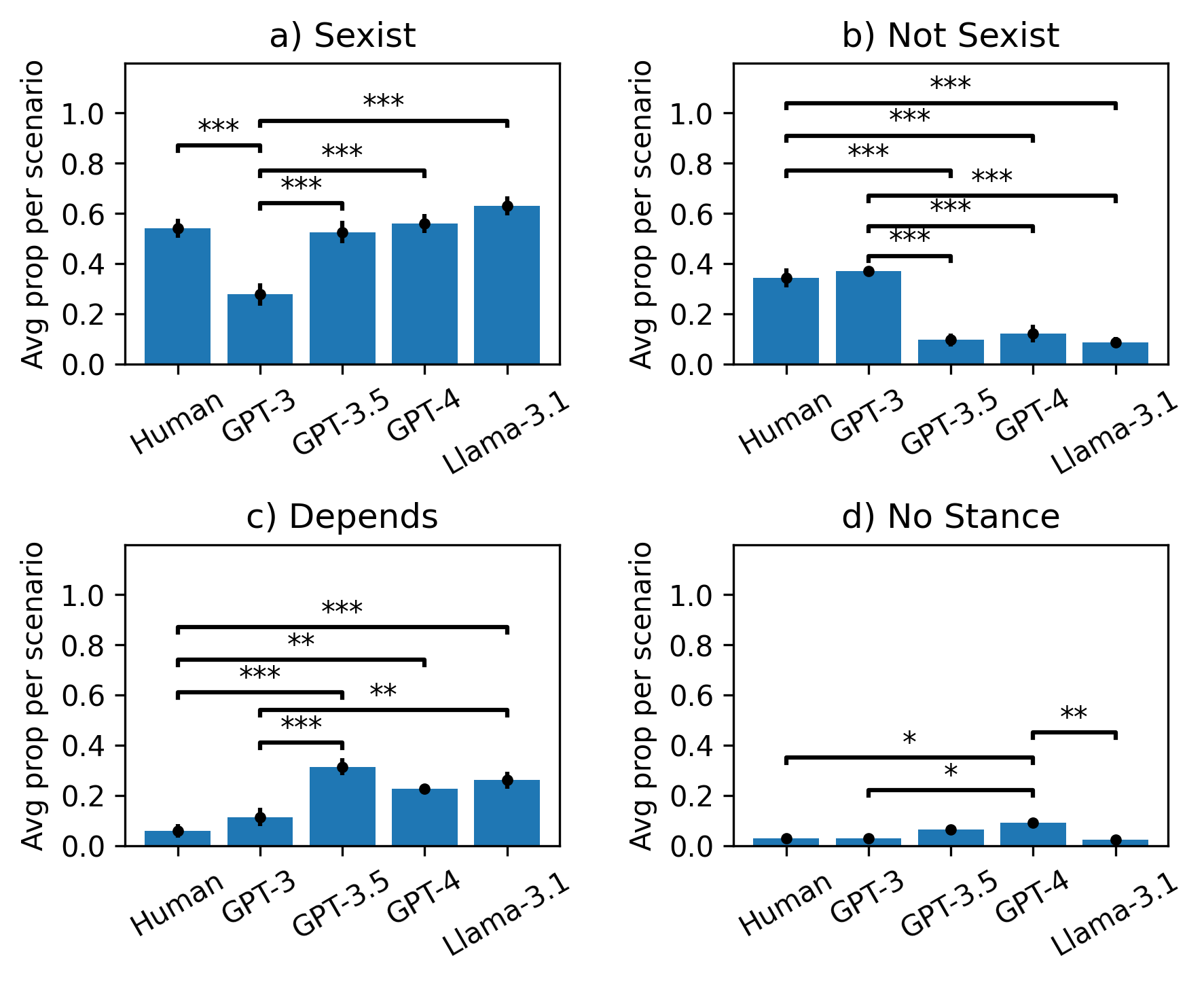}
    \caption{Average proportion of each scenario's explanations that contains each Stance, per dataset. *** represents $p<.001$, ** $p<.01$, * $p<.05$. Error bars represent one standard error}
    \label{fig:StancesChart_dec04}
    \Description{Four bar plots displaying the average proportion of responses per scenario, averaged across scenarios, which display each Stance. The x-axis shows the authors of responses: human, GPT-3, GPT-3.5, GPT-4 and Llama 3.1. The top left plot shows statistical significance in the frequency of the Stance \emph{sexist}, where it is more common in the human dataset, GPT-3.4, GPT-4 and Llama 3.1 than on GPT-3. The top right plot shows the Stance \emph{not sexist}, and how it is more common in GPT-3 and human dataset than GPT-3.5, GPT-4 or Llama 3.1. The bottom left shows the \emph{depends} Stance, where the human dataset, GPT-3.5 and GPT-4 display this Stance more often than GPT-3 and Llama 3.1. The bottom right displays \emph{no stance}, where there is more in GPT-4.}
\end{figure*}

\emph{Stance} relates to the classification of scenarios in one of the following categories: \emph{sexist}, \emph{not sexist}, \emph{depends} or \emph{no stance} as defined in Table \ref{tab:CodingScheme_Stance}.
We first looked into raw frequencies and the average of scenario-level aggregations, as shown in Table \ref{tab:stance-total}. Scenario-level aggregations were required since each dataset had a different number of entries once non-coherent entries were filtered out, as described in \textit{Methods}. 
We then compared the frequencies of each \textit{Stance} between the datasets using a four-factor ANOVA with Bonferroni post-hoc tests. These tests revealed significant differences in all four Stance categories, as seen in Figure \ref{fig:StancesChart_dec04}.

\renewcommand{\arraystretch}{1.5}
\begin{table*}[]
\centering
\caption{Frequency of Stances present in human and LLM-generated responses of subtle sexism scenarios. Note that each response could only be coded as one stance. The ``Average of scenario-level aggregations" refers to the average proportion of responses containing each Stance when aggregated at the scenario level, and then averaged across all scenarios.}
\resizebox{\textwidth}{!}{%
\label{tab:stance-total}
\begin{tabular}{lllllllllll}
                    & \multicolumn{5}{l}{\textbf{Raw Frequencies}}                                                                                                                                                                                                                                                                               & \multicolumn{5}{l}{\textbf{Average of scenario-level aggregations}}                                                                                                                                                                                                                                   \\ \cline{2-11} 
                    & \begin{tabular}[c]{@{}l@{}}Human\\ (n=307)\end{tabular} & \begin{tabular}[c]{@{}l@{}}GPT-3\\ (n=229)\end{tabular} & \begin{tabular}[c]{@{}l@{}}GPT-3.5\\ (n=303)\end{tabular} & \begin{tabular}[c]{@{}l@{}}GPT-4\\ (n=303)\end{tabular} & \multicolumn{1}{l|}{\begin{tabular}[c]{@{}l@{}}Llama 3.1\\ (n=303)\end{tabular}} & \begin{tabular}[c]{@{}l@{}}Human\\ (n=307)\end{tabular} & \begin{tabular}[c]{@{}l@{}}GPT-3\\ (n=229)\end{tabular} & \begin{tabular}[c]{@{}l@{}}GPT-3.5\\ (n=303)\end{tabular} & \begin{tabular}[c]{@{}l@{}}GPT-4\\ (n=303)\end{tabular} & \begin{tabular}[c]{@{}l@{}}Llama 3.1\\ (n=303)\end{tabular} \\ \hline
\textit{Sexist}     & 163 (53\%)                                              & 83 (36\%)                                               & 159 (52\%)                                                & 170 (56\%)                                              & \multicolumn{1}{l|}{191 (63\%)}                                                  & 54\%                                                    & 25\%                                                    & 52\%                                                      & 56\%                                                    & 62\%                                                        \\
\textit{Not Sexist} & 116 (38\%)                                              & 112 (49\%)                                              & 28 (9\%)                                                  & 37 (12\%)                                               & \multicolumn{1}{l|}{26 (9\%)}                                                    & 34\%                                                    & 37\%                                                    & 9\%                                                      & 12\%                                                    & 7\%                                                        \\
\textit{Depends}    & 18 (6\%)                                                & 26 (11\%)                                               & 96 (32\%)                                                 & 68 (22\%)                                               & \multicolumn{1}{l|}{79 (26\%)}                                                   & 5\%                                                    & 11\%                                                    & 31\%                                                      & 23\%                                                    & 31\%                                                        \\
\textit{No Stance}  & 10 (3\%)                                                & 8 (3\%)                                                 & 20 (6\%)                                                  & 28 (9\%)                                                & \multicolumn{1}{l|}{7 (2\%)}                                                     & 2\%                                                    & 3\%                                                    & 6\%                                                      & 9\%                                                    & 2\%                                                        \\ \hline
\end{tabular}
}
\end{table*}

\subsubsection{Stance: sexist} 
The Llama 3.1 dataset had the highest count of \emph{sexist} stance in raw frequencies and average scenario-level aggregations. In contrast, GPT-3 outputs used this classification
significantly less often than all other datasets ($p<.001$ for all). A shared trait in human and LLM responses (including GPT-3) when justifying that a scenario is \emph{sexist} was the use of `it implies', `it generalizes, `it stereotypes' (and other variations) where `it' refers to the scenario, action or behaviour being assessed: \textit{``This is sexist because \textbf{it clearly assumes} women can't be courageous or strong"} (Human). Similarly, cultural gender expectations were a common denominator in explanations with \emph{Sexist} stance: \textit{``Yes, the statement can be considered sexist. It \textbf{perpetuates gender stereotypes} by suggesting that men are better at parking and enjoy challenges (...)"} (GPT-3.5). 


\subsubsection{Stance: not sexist}
We also found significant differences for the occurrences of \emph{not sexist} stance ($F(4, 488)=24.51, MSE = 0.0791, p<.001$), as it was found more frequently in GPT-3 than in GPT-3.5, GPT-4 and Llama 3.1 outputs (all at $p<.001$), but there was
no statistically significant difference when comparing GPT-3 and the human dataset. Interestingly, both humans and GPT-3 displayed the \emph{not sexist} stance often in short and definitive explanations that argue something cannot be sexist if it is true: \textit{``No, no, no. It's not sexist. It's true."} (GPT-3); and \textit{``hell no because it's true''} (Human). Thematically, it was common across all datasets to point out the behaviour was impolite, but \emph{not sexist}: \textit{``Not sexist. People are just \textbf{rude} sometimes."} (Human). Similarly: \textit{"(...) It doesn't clearly show a prejudice or discrimination based on gender. However, it could be seen \textbf{as rude or inconsiderate} to ignore your partner, but that's more about manners and professional etiquette than sexism"} (GPT-4). 

\subsubsection{Stance: depends}
This Stance differed across human and model datasets ($F(4, 488)=11.78, MSE=0.0926, p<.001$) with GPT-3.5, GPT-4 and Llama 3.1 displaying it much more often, perhaps highlighting that newer model outputs contain more nuance. The \emph{depends} stance is characterized by conditions that would change the final assessment, for example: \textit{``(...) Ultimately, whether being chivalrous is sexist \textbf{depends} on the context, intentions, and power dynamics involved (...)"} (Llama 3.1). The prevalence of this Stance in the newer models is also characterized by thoroughly considering the scenario from multiple angles and demonstrating cautious deliberation. \textit{``The word `hysterical' itself is \textbf{not inherently sexist}. It is generally used to describe someone who is behaving or acting in an extremely emotional and uncontrolled way. \textbf{However}, it's important to note that its historical usage has sexist undertones. The term originates from the Greek word "hyster" meaning uterus, and it was thought that hysteria was a disease only of women and caused by disturbances in the uterus. Using "hysterical" to belittle women's emotions \textbf{can thus be seen as sexist} when it perpetuates harmful stereotypes about gender and emotion. \textbf{It depends on the context} it's used in." (GPT-4)}.

\subsubsection{Stance: no stance}
GPT-4 had the most outputs with \emph{no stance} when no final decision was made in regards to Stance. We found significant differences between datasets
($F(4, 488)=4.043, MSE=0.02182, p<.01$), 
with the most significant difference between GPT-4 and Llama 3.1 ($p<0.01$). We also found significant differences between humans and GPT-4, as well as between GPT-3 and GPT-4 (both $p<0.05$). This stance was commonly marked by personal uncertainty, (i.e. `on the fence'). First-person pronouns were found in GPT-3 (e.g.,\textit{``\textbf{I can't really say}"} (GPT-3) and in the human dataset (e.g., \textit{``(...) Is affirmative action discrimination? Is discrimination objective or subjective?   \textbf{I am not sure}."}). In contrast, GPT-3.5, GPT-4, and Llama 3.1 adopted a more impersonal approach, asking for more context and information to make the final judgment:`\textit{``(...) it's \textbf{hard to definitively label} this situation as sexist."} (GPT-4).


\subsubsection{Consistency in Stance}
We also analyzed the consistency between responses to the same scenario (i.e., when different human participants responded to the same scenario, did they opt for the same \emph
Stance? And for LLMs, as each model was prompted three times, were the three responses consistent in the Stance assessment?). Because the codes in \emph{Stance} are mutually exclusive (i.e., one scenario could only be classified using one code), we calculated the average agreement per scenario, per dataset, and found that, on average, 78\% of human responses attributed the same Stance when assessing the same scenario (73\% in GPT-3, 84\% in GPT-3.5, 84\% in GPT-4 and Llama 3.1 83\%). This means that, across the multiple human responses for each scenario, 78\% of the human participants argued for the same Stance, on average. For the three GPT-3.5 responses per scenario, the same stance was used 84\% of the time. Then we calculated the \% of all scenarios with 100\% consistency in responses per scenario: human dataset had 100\% consistency in 40\%, GPT-3 37\%, GPT-3.5 56\%, GPT-4 58\%, Llama 55\%. Thus, humans and models were relatively consistent on a scenario level---with GPT-3 being the least consistent overall and GPT-3.5 and GPT-4 being the most consistent in Stance assessment.

\subsection{Perspectives}

Similarly to Stance, we first analyzed the raw frequencies of each Perspective per dataset (Table \ref{tab:perspective-total}), followed by a four-factor ANOVA with Bonferroni post-hoc tests. Note that some responses did not contain any Perspectives (i.e., the author did not consider any point of view to justify rationale) or combined multiple Perspectives. Thus, the total of Perspectives is not expected to reach 100\%.

\renewcommand{\arraystretch}{1.5}
\begin{table*}[]
\centering
\caption{Frequency of Perspectives present in human and LLM-generated text. Note that each explanation could contain multiple or no Perspectives, hence values will not sum to 100. Includes proportion averages when aggregated at the scenario level, and then averaged across all scenarios.}
\label{tab:perspective-total}
\resizebox{\textwidth}{!}{%
\begin{tabular}{llllll|lllll}
                        & \multicolumn{5}{c|}{\textbf{Raw Frequencies}}                                                                                                                                                                                                                                                         & \multicolumn{5}{c}{\textbf{Average of scenario-level aggregations}}                                                                                                                                                                                                                                   \\ \cline{2-11} 
                        & \begin{tabular}[c]{@{}l@{}}Human\\ (n=307)\end{tabular} & \begin{tabular}[c]{@{}l@{}}GPT-3\\ (n-229)\end{tabular} & \begin{tabular}[c]{@{}l@{}}GPT-3.5\\ (n=303)\end{tabular} & \begin{tabular}[c]{@{}l@{}}GPT-4\\ (n=303)\end{tabular} & \begin{tabular}[c]{@{}l@{}}Llama 3.1\\ (n-303)\end{tabular} & \begin{tabular}[c]{@{}l@{}}Human\\ (n=307)\end{tabular} & \begin{tabular}[c]{@{}l@{}}GPT-3\\ (n-229)\end{tabular} & \begin{tabular}[c]{@{}l@{}}GPT-3.5\\ (n=303)\end{tabular} & \begin{tabular}[c]{@{}l@{}}GPT-4\\ (n=303)\end{tabular} & \begin{tabular}[c]{@{}l@{}}Llama 3.1\\ (n=303)\end{tabular} \\ \hline
\textit{Decision-maker} & 44 (14\%)                                               & 87 (38\%)                                               & 34 (11\%)                                                 & 5 (2\%)                                                 & 13 (4\%)                                                    & 12\%                                                    & 40\%                                                    & 11\%                                                      & 2\%                                                    & 4\%                                                        \\
\textit{Victim}         & 45 (15\%)                                               & 59 (26\%)                                               & 256 (84\%)                                                & 249 (82\%)                                              & 288 (95\%)                                                  & 16\%                                                    & 24\%                                                    & 85\%                                                      & 82\%                                                    & 95\%                                                        \\
\textit{Perpetrator}    & 85 (28\%)                                               & 80 (35\%)                                               & 103 (34\%)                                                & 131 (43\%)                                              & 125 (41\%)                                                  & 26\%                                                    & 35\%                                                    & 34\%                                                      & 43\%                                                    & 41\%                                                        \\ \hline
\end{tabular}
}
\end{table*}

\subsubsection{Decision-maker Perspective} This Perspective is characterized by the use of personal pronouns (I, me, mine) explicitly marking the author's presence in the response. GPT-3 had the highest raw frequency, found in 38\% of this model's responses. The frequency of the \emph{decision-maker's} perspective was significantly different across data sets ($F(4, 492)=48.34, MSE =0.0491, p<.001$), with GPT-3 being significantly different from all other models and humans (at $p<.001$ in all comparisons). A smaller but still significant difference was found between GPT-4 and the human dataset ($p = 0.01$). For the \emph{decision-maker} Perspective, humans used personal experiences as part of their rationale (e.g., \textit{``maybe. back in the day the person saying that would typically be a male, \textbf{I heard it a lot}."} (Human); a similar strategy was found in the GPT-3 dataset, where the model mimicked human-like personal preferences and experiences: \textit{``(...) What I can say is that \textbf{in my experience}, it is a sign of warmth and friendship. (...)  \textbf{I have kissed men on the cheek.} (...)"} (GPT-3). In contrast, GPT-3.5 and GPT-4 responses used the \emph{decision-maker}'s Perspective to state the model's limitations, suggesting built-in guardrails for newer models, for example: \textit{``As an \textbf{AI language model, I don't have personal opinions or feelings}, but I can provide an analysis of the statement you shared."} (GPT-3.5); and \textit{``As \textbf{an AI, I don't have personal beliefs or emotions}, but I can provide analysis on this issue."} (GPT-4). Notably, the most recent and open-source model, Llama 3.1, did not have similar guardrails, and the use of personal pronouns did not indicate either if the author is human or AI: \textit{``After analyzing the text, \textbf{I wouldn't} classify the situation as sexist."} (Llama 3.1).

\subsubsection{Victim's Perspective}
The \emph{victim’s} perspective becomes much more prevalent in the newer GPT models and Llama 3.1, differing across datasets ($F(4, 492)=204, MSE =0.067, p<.001$) while having significant differences between almost all model and human combinations--- except for humans and GPT-3 as well as GPT-3.5 and GPT-4 where differences were not significant. Remarkably, Llama 3.1 had \emph{victim} Perspective on 95\% of responses: \textit{``The statement reduces women to objects, implying that their bodies are mere triggers for men's thoughts and desires. This objectification ignores women's agency, autonomy, and individuality".} (LLama 3.1). On the other hand, GPT-3 used victim outcomes less frequently than all other LLMs (all at $p<.001$); however, the significance was not found between GPT-3 and the human dataset. Thematically, the \emph{victim's} Perspective was similar across all datasets. It was either used to highlight negative outcomes to an individual in the scenario, for example: \textit{``Yes this is they took advantage of \textbf{her} work ethic and made a joke out of it to make \textbf{her} feel inferior."} (Human). Or it was used to show that the offence caused harm to a group, as in:  \textit{`` It's sexist to assume that \textbf{men} cannot be good babysitters"} (GPT-3). Expectations based on gender were also common, for example, comments about physical capabilities (e.g., \textit{``(...) it's saying women are inherently \textbf{weaker} than men, which implies they can't handle certain tasks or a certain amount of physical labor"} (Human)), physical appearance (e.g., \textit{``It suggests that society may be more accepting of men's bodies, specifically dad bods, while potentially disregarding or not appreciating the \textbf{appearance} of mothers or women in general."} (GPT-3.5)), or roles in society (e.g., \textit{``(...) This type of language can perpetuate a \textbf{power imbalance and reinforce stereotypes} about men being the ones in charge and women being the ones who do domestic tasks. (...)"} (Llama 3.1)).

\subsubsection{Perpetrator's Perspective} GPT-4 had the highest raw frequency and average of scenario-level aggregations for this perspective, leveraging the beliefs and intentions of a person or group accused of sexism into account, such as in \textit{``This situation might be indicative of a lack of awareness or consideration on the part of \textbf{the boyfriend}, but it doesn't necessarily mean he is behaving this way because of gender-based bias or prejudice."} (GPT-4). Overall, the significance was weaker compared to other Perspectives ($F(4, 492)=3.869, MSE=0.1205, p=0.01$), and driven by a higher frequency in GPT-4 ($p<0.01$) and Llama 3.1 ($p<0.05$), in comparison to the human dataset. Thematically, all datasets had responses pointing out that the perpetrator \emph{did not} intend to be sexist or cause harm: \textit{``It's worth noting that the assistant's behaviour could also be due to a lack of awareness or training on inclusive customer service, rather than any intentional sexism."} (Llama 3.1). Attempts at humour or irony were another common justification: \textit{``This is \textbf{humorous self-depreciation}. It’s not aimed at women."} (Human). Furthermore, this Perspective was used to `shift the blame' from the individual to external influences such as family or societal context: \textit{``Traditional, not necessarily sexist.  Women often chase men out of the kitchen \& tell them to go watch the television or something."} (Human). Conversely, the perpetrator's intentions could also be interpreted as being positive towards the victim, in what is known as \emph{benevolent sexism} (e.g., chivalry) \cite{hammond2018benevolent}: \textit{`` No it is not sexist. When a man offers to help, you should take him up on his offer. It is \textbf{chivalrous, gentlemanly}. It makes you feel like a woman."} (GPT-3). Interestingly, while human responses often brought a single interpretation about the perpetrator's intentions (e.g., they meant to cause harm), GPT-3.5 and GPT-4 speculated multiple interpretations of intent: \textit{``(...) If someone suggested that you should open a Twitch profile so people can watch you play video games because they believe you are good at gaming or they think you're entertaining, it is not sexist. However, if they suggested it because they believe women should or shouldn't do certain activities due to stereotypical gender roles, then it may be considered sexist. (...)"} (GPT-4).

\begin{figure*}
    \centering
    \includegraphics[width=0.75\linewidth]{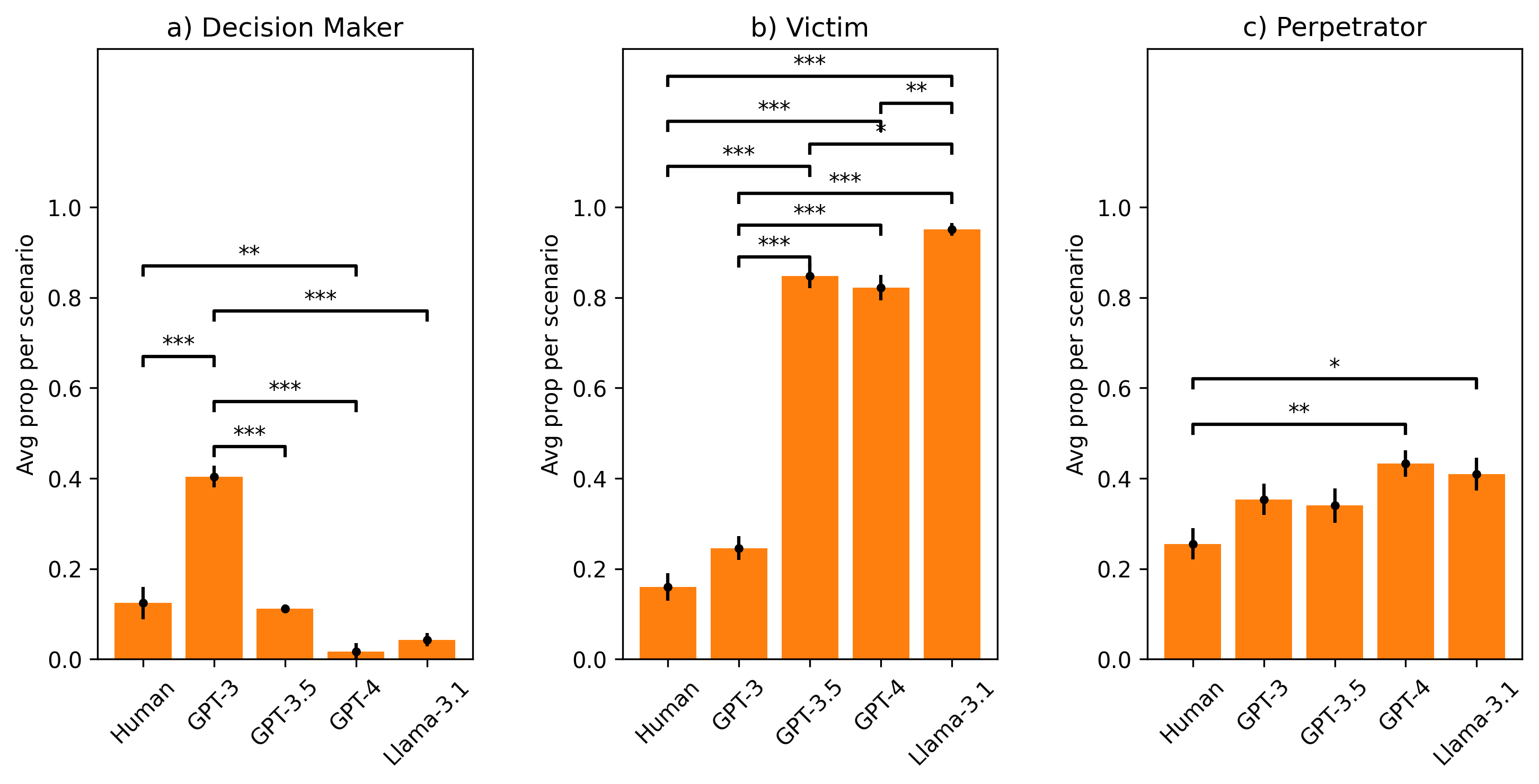}
    \caption{Average proportion of each scenario's responses containing each Perspective, per dataset. *** represents $p<.001$, ** $p<.01$, * $p<.05$. Error bars represent one standard error.}
\Description{Four bar plots displaying the average proportion of responses per scenario, averaged across scenarios, which display each Perspective. The x-axis contains the four datasets studied, humans, GPT-3, GPT-3.5, and GPT-4. The leftmost bar plot displays the \emph{decision-maker} Perspective, showing that it is much more common in GPT-3 than any other dataset and is more common in human and GPT-3.5 than GPT-4. The middle plot refers to the \emph{victim's} Perspective and shows that this Perspective is much more common in GPT-3.5 and GPT-4 than GPT-3 or humans. The right-most plot focuses on the \emph{perpetrator} Perspective and shows that this is only more common in GPT-4 than in human.}
    \label{perspectives_updated}
\end{figure*}

\subsubsection{Combined Perspectives}
Given that some responses contained multiple Perspectives (e.g., a response considered both \emph{victim} outcomes and \emph{perpetrator} intentions), we counted when this occurred for each dataset (see Table \ref{tab:table 5} for mean and standard deviation). The human dataset had the lowest count of perspectives per response, with human participants opting for combined perspectives in 17 out of 307 responses (6\%). For LLMs, the findings suggest combinations become increasingly more common in the newer model iterations: GPT-3 (26\% of the scenarios had combinations; mean=0.99; STD=0.75), GPT-3.5 (29\%; mean=1.3; STD=0.64), GPT-4 (30\%; mean=1.27; STD=0.54), Llama 3.1 (39\%; mean=1.40; STD=0.59).

\begin{table}
    \centering
    \caption{Number of Perspectives shown per response for each dataset. STD = standard deviation}
    \begin{tabular}{lll}
      Dataset   &  Mean  & STD\\\toprule
       Human  & 0.57 & 0.60\\
        GPT-3 & 0.99 & 0.75\\
        GPT-3.5 & 1.30  & 0.64\\
        GPT-4 &  1.27 & 0.54 \\
        Llama 3.1 & 1.40 & 0.59 \\\bottomrule
    \end{tabular}
    
    \label{tab:table 5}
\end{table}

We then analyzed which Perspectives were being combined (Table \ref{tab:multiple-perspectives}). The most common combination in human and GPT-3 datasets was \emph{decision-maker and perpetrator}, while this was very rare in GPT-3.5 and GPT-4, and not present in Llama 3.1. On the other hand, GPT-3.5, GPT-4 and Llama 3.1 had \emph{victim and perpetrator} bundled together more often than human and GPT-3 datasets. This aligns with the fact that the \emph{victim} Perspective is used much more often in these newer models than in humans or GPT-3, while the \emph{perpetrator} Perspective is used approximately equally. This observation suggests that the \emph{victim} Perspective is commonly added by newer models to explanations where the \emph{perpetrator} perspective is already present. GPT-3 and GPT-3.5 also sometimes combine the \emph{decision-maker and victim} Perspectives, which was uncommon in other datasets. The combination of all Perspectives was rare and occurred mostly in GPT-3.5 and Llama 3.1.

\begin{table*}    
\centering
\caption{Combination of Perspectives per dataset and respective percentages. Note some responses contained only one Perspective, hence the sum is not expected reach 100\%.}
\label{tab:multiple-perspectives}
\begin{tabular}{llllll}   
Combined Perspectives                      & \begin{tabular}[c]{@{}l@{}}Human \\ (n=307)\end{tabular} & \begin{tabular}[c]{@{}l@{}}GPT-3 \\ (n=229)\end{tabular} & \begin{tabular}[c]{@{}l@{}}GPT-3.5\\ (n=303)\end{tabular} & \begin{tabular}[c]{@{}l@{}}GPT-4\\ (n=303)\end{tabular} & \begin{tabular}[c]{@{}l@{}}Llama 3.1\\ (n=303)\end{tabular} \\ \hline
Decision-maker \& Victim                & 1 (\textless{}1\%)                                       & 15 (7\%)                                                 & 12 (4\%)                                                  & 2 (1\%)                                                 & 2 (1\%)                                                     \\
Decision-maker \& Perpetrator           & 12 (4\%)                                                 & 32 (14\%)                                                & 3 (1\%)                                                   & 1 (\textless{}1\%)                                      & 0                                                           \\
Victim \& Perpetrator                   & 3 (1\%)                                                  & 9 (4\%)                                                  & 55 (18\%)                                                 & 86 (28\%)                                               & 107 (35\%)                                                  \\
Decision-maker \& Victim \& Perpetrator & 1 (\textless{}1\%)                                       & 2 (1\%)                                                  & 17 (6\%)                                                  & 2 (1\%)                                                 & 10 (3\%)                                                    \\ \hline
Total responses with combinations       & 17 (6\%)                                                 & 58 (26\%)                                                & 87 (29\%)                                                 & 91 (30\%)                                               & 119 (39\%)                                                 
\end{tabular}
\end{table*}

\subsubsection{Consistency in Perspectives}
To calculate consistency for \emph{Perspectives}, the codes were not mutually exclusive (i.e., one response could have more than one Perspective), therefore we calculated the Jaccard Similarity \cite{eagan2020testing} for each scenario, and then averaged across all scenarios. This would be equivalent to using Jaccard similarity to measure inter-rater reliability, where each human or LLM output acts as a rater. Our findings indicate that the human and GPT-3 datasets had similar, although low, consistency when assigning Perspectives with an average Jaccard similarity across scenarios of 0.29. Additionally, the results point to a progression in consistency with newer model iterations: GPT-3 (0.29), GPT-3.5 (0.57), GPT-4 (0.59) and Llama 3.1 (0.66).

\subsection{Stance x Perspective}
To understand the interaction effects between \emph{Stance} and \emph{Perspectives}, we started with a frequency count. For instance, across all datasets, when the scenarios were classified as \emph{sexist}, the \emph{victim}'s perspective was the most prevalent (human = 36, GPT-3 = 39, GPT-3.5 = 161, GPT-4 = 161, Llama 3.1 = 188). In \emph{not sexist}, human, GPT-3.5 and GPT-4 datasets had \emph{perpetrator} most often (human = 45, GPT-3.5 = 14, GPT-4 = 21) while GPT-3 had \emph{decision-maker} (GPT-3 = 42) and Llama 3.1 had \emph{victim} (Llama 3.1 = 19). The last two results might look surprising (e.g., Llama 3.1 responses to \emph{not sexist} containing the \emph{victim}'s Perspective more often), but show that there is little correlation between Stance and Perspective, above and beyond the patterns in distributions of Perspectives shown in  Table \ref{tab:perspective-total}. In \emph{Depends}, human, GPT-3 and GPT-4 opted more for \emph{perpetrator}, while GPT-3.5 and Llama 3.1 had \emph{victim}. Lastly, in \emph{no stance}, human and GPT-3 were aligned displaying the \emph{decision-maker}'s perspective, while GPT-3.5, GPT-4 and Llama 3.1 tended towards \emph{victim}.
Overall, the results align with the Perspective that was most common in the dataset (e.g., Llama 3.1 had the most counts of \emph{victim}, thus it was prevalent across all Stances).

We also looked into how \emph{Stance} was defined when multiple \emph{Perspectives} were present in the output. In the human dataset, the most common combination had \emph{decision-maker and perpetrator} and these responses were often classified as \emph{not sexist}. However, in a few instances, the \emph{decision-maker} considered the scenario was \emph{sexist} despite the \emph{perpetrator}'s intentions \textit{``Yes, but indo (sic) not think it is intentional. I think that societal norms have created an idea of the roles men and women should play, and for women to step into and succeed in a male role that is very clever."} (Human). In the \emph{victim and perpetrator} combination, most common in the newer models where more nuance is displayed, GPT-3.5 often classified the scenario as \emph{depends}, GPT-4 balanced between \emph{sexist} and \emph{depends}, and Llama 3.1 mostly \emph{depends}. Lastly, when combining all three Perspectives, GPT-3.5 had the most occurrences, and in such instances, the scenarios were classified as \emph{depends} and \emph{no stance}. Here is an example of the latter, where GPT-3.5's response considers multiple angles, explicitly names model limitations and expresses that a final decision is not possible: \textit{``\textbf{As an AI language model, I can provide} an analysis but \textbf{cannot provide a definitive answer} as to whether something is sexist or not. (...) Some might argue that the lyrics \textbf{reinforce traditional gender roles and stereotypes, presenting the woman as subservient} and the man as dominant. Others might interpret the lyrics differently, seeing them \textbf{as merely describing a romantic relationship} without necessarily implying sexism."} (GPT-3.5).

Next, we were interested in examining if LLM outputs displayed perspectives that human responses \emph{did not} when assessing \emph{the same scenario}. Using frequency count, we identified GPT-3 had one or more perspectives humans did not in 65\% of the scenarios; GPT-3.5 (72\%), GPT-4 (71\%), Llama (75\%). In GPT-3, the complementary perspective that was not present in human responses was most often the \emph{decision-maker}'s, aligned with the finding that this model had the most counts of this perspective (see Table \ref{tab:perspective-total}). On the other hand, GPT-3.5, GPT-4 and Llama 3.1 had the \emph{victim}'s perspective as most prevalent in this analysis.

\subsection{Results Summary} Our comparative analysis using quantitative and qualitative methodologies revealed significant differences in responses produced by humans and LLMs (GPT-3, GPT-3.5, GPT-4, and Llama 3.1) when prompted to assess subtle sexism scenarios.

The GPT-3 dataset had significantly more  \emph{not sexist} Stance and \emph{decision-maker} Perspective. The GPT-3.5 and GPT-4 datasets were similar to Llama 3.1 in \emph{depends} Stance being used frequently, considering the scenario from multiple angles. Distinctively, GPT-4 had the highest frequencies of \emph{no stance} and \emph{perpetrator} Perspective compared to all models and human datasets. Additionally, Llama 3.1 had the highest count of the \emph{sexist} Stance across the board and 95\% of responses contained the \emph{victim} Perspective for justification.

Interestingly, the GPT-3 dataset had the most similarities to the human dataset in a few criteria. First, both had low consistency in their assessments for Stance and Perspectives, indicating higher variability in outputs when compared to GPT-3.5, GPT-4 and Llama 3.1. Second, humans and GPT-3 also presented personal experiences and opinions in their argumentation, with GPT-3, in particular, displaying the highest count of \emph{decision-maker} of all datasets. In fact, the GPT-3 dataset had some responses mimicking human experiences at times indicating the model is self-identifying as a woman (e.g., \textit{``(...) \textbf{As a woman, I am allowed to be girly}, just like men are allowed to be masculine. (...)"} (GPT-3)) which could lead to anthropomorphism and/or cause the uncanny valley effect \cite{mori2012uncanny}. In contrast, Perspectives was also a useful parameter to identify AI safety guardrails \cite{vicente2023humansinherit} enforced in the newer GPT models (GPT-3.5 and GPT-4) as they clearly state their limitations as an AI model. Llama's responses did not display the same strategy but curiously often contained unsolicited advice to rephrase the scenario to make it ``less sexist" or `'less offensive", even though the prompt did not have instructions for such, as exemplified in \textit{"\textbf{If I had to suggest} a rephrased version that still conveys the same message without leaning on sexist assumptions, I might recommend (...)"} (Llama 3.1).

Our results also indicate the human dataset had the lowest count of Perspectives and combinations of Perspectives per response, indicating the human participants offered more simplistic explanations as they did not consider multiple points of view to inform their assessment. When comparing to LLMs in this aspect, there is a progressive increase in combined perspectives, as GPT-3 the, older GPT series models in our study (released in 2020) had the lowest count of perspectives and perspective combinations, as opposed to Llama 3.1, (released in 2024), more details in Tables \ref{tab:table 5} and \ref{tab:multiple-perspectives}. This is further evidenced by GPT-3.5 and GPT-4 outputs with high counts of \emph{depends} and \emph{no stance}, indicating cautiousness in assessments, either by considering the factors that would influence a decision or by deliberately not taking a stance. Furthermore, our analysis suggests the newer the model, the more chances of perspectives being displayed that were not present in human responses to the same scenario, thus indicating the potential for complementarity. When there was a disagreement in \emph{Stance} for the same scenario, GPT-3.5, GPT-4 and Llama 3.1 outputs most often offered the \emph{victim}'s Perspective when the human participants did not.

\section{Discussion}
In subjective human-AI collaboration, when the scenario is open to interpretation, the complementarity of AI input \cite{hemmer2021human, bansal2021does} cannot be measured by typical objective parameters such as accuracy, robustness, or consistency \cite{liang2023HELM,fok2023search}. Instead, we suggest that the effectiveness of AI input can be assessed through the complementarity of perspectives, i.e. points of view, provided by AI and considered by the human. In particular, our work has outlined two sets of parameters that can be useful to observe in subjective decision-making: \emph{Stances} and \emph{Perspectives}. Our results demonstrate humans and LLMs (GPT-3, GPT-3.5, GPT-4 and Llama 3.1) consistently use the same parameters but at different frequencies and in various combinations, thus indicating each model displayed tendencies to adopt and combine \emph{Stances} and \emph{Perspectives}. For example, outputs from the most recent LLMs (GPT-3.5, GPT-4 and Llama 3.1) added and combined more \emph{Perspectives} when compared to GPT-3 and to human responses, indicating a tendency towards more nuanced assessments, considering multiple points of view in the scenario. In contrast, the human responses had more simplistic argumentation with less diversity in \emph{Perspectives}. In this section, we discuss the implications of these findings to the HCI field.

\subsection{Diversity of Stances and Perspectives}

Arguably, the need for diversity in \emph{Stances} and \emph{Perspectives} displayed by a model can be determined by the use case and differ correspondingly. Some instances might require more tempered assessments such as displayed when models classified scenarios as \emph{depends} or \emph{no stance} (found more often in GPT-3.5 and GPT-4 but not as often in Llama 3.1). In practice, for example, if a model is playing the role of devil's advocate \cite{chiang2024devil}, its usefulness will depend on bringing arguments from unique and opposing \emph{Perspectives} to foster thoughtful debate. Other examples include using AI to produce counterspeech \cite{counterspeakers} or to facilitate conflict resolution \cite{Conflict.resolution}. This aligns with recent discussions in HCI that suggest exploring different roles AI can play in human-AI collaboration for decision-making processes, beyond the traditional AI recommender systems where the ultimate goal is to provide the single most accurate output \cite{ma2024beyond}. Due to recent concerns of overreliance on AI recommendations and calls for trust calibration mechanisms \cite{zhang2020effect, wischnewski2023measuring}, one popular approach suggests it might be appropriate, in some use cases, for AI to offer the pros and cons \cite{miller2023explainable} or pose questions \cite{danry2023don} that will stimulate critical thinking \cite{Sarkarchallenge} and potentially lead to appropriate reliance levels \cite{chiang2024devil}. To design AI systems to fulfill this type of role, \emph{Stance} and \emph{Perspectives} can be applied as parameters in fine-tuning \cite{howard2018universal}, especially when contextual interpretation of nuanced social interactions is required. For instance, in industry, a GPT-3.5-powered chatbot named \emph{Vera} \cite{Fermin_2023} was designed to advise human resource professionals and promises to \textit{``help you through a complex employee relations issue"}. In dialogue with the user, \emph{Vera} interprets a work-related situation and provides recommendations for the next steps. When dealing with interpersonal conflict at the workplace, contemplating the different points of view of actors in the scenario is crucial, and \emph{Perspectives} could be used to define which ones, at what frequency and how to surface them -- i.e., parameters to fine-tune pre-trained models such as the GPT series. Similarly, \emph{Stance} could be used to calibrate or evaluate the decisiveness of a model's outputs. For instance, the \emph{depends} Stance, most present in the GPT-3.5 dataset, can be useful when the decision-maker needs to pinpoint the critical factor (e.g., intentions behind the action, social context, etc.) that would `tip the scales' and make a scenario be deemed sexist or not. Alternatively, \emph{no stance}, most found in GPT-4 dataset, could be leveraged when there is a need to verify how often a model discloses their limitations, as the response clearly states when more information or context is needed before a final decision is made. Furthermore, as our analysis demonstrates, the newer LLM model responses had more comprehensive and methodical analyses of a scenario, combining different Perspectives whereas, in contrast, human rationale tended to be rich in personal beliefs and experiences. 

It should be noted that the multitude of potential interpretations often leads to challenges for annotator agreement in label classification 
when applied to non ground truth tasks \cite{dutta2023modeling, akhtar2020modeling}, such as content moderation \cite{garg2023handling}. For instance, past work \cite{waseem-hovy-2016-hateful, waseem2016you} reported low agreement when human annotators classify hate speech datasets and confirmed the diversity of moral values impacts what is classified as offensive language \cite{disentangling.offensiveness1}. Efforts to remove ambiguity in annotated datasets often employed the ``majority vote" \cite{atallah2019heart} (i.e., choose the label with the most votes), which was criticized by researchers \cite{lease2011quality, garg2023handling} for disregarding nuances and potentially introducing biases. Alternatively, \citet{gordon2022jury} suggests multiple annotator perspectives could be leveraged especially in these subjective decision-making tasks, provided all voices are balanced for fairness \cite{garg2023handling}. This experiment aligns with recent developments in the \emph{Perspectivist Approach} arguing that diverse perspectives should be included in detriment to single ground truth proxies \cite{baron2014groundless}. Comparably, \citet{sorensen2024roadmap} suggests the adoption of 
\emph{Pluralistic alignment}, when LLMs are faced with open-to-interpretation questions and multiple relevant responses are possible, arguing this improves fairness, aiming for systems that are \textit{``capable of representing a diverse set of human values and perspectives"}. 

\subsection{Implications for LLM Evaluation}
Given that the multiplicity of perspectives is an inherent characteristic of subjective decision-making tasks such as assessing subtle sexism, we suggest that designing an LLM-powered system to assist in subjective decisions requires an ability to evaluate this multiplicity in the corresponding model. LLM evaluations are typically performed via automated evaluation with automatically computed evaluation criteria \cite{liang2023HELM}, human evaluation \cite{chiang2024chatbot}, or a combination of the two  \cite{awsbedrock}. 
Since human evaluation can be costly to operationalize, automated evaluation is often preferred. However, past work has criticized automated evaluation techniques for their fixed nature and for missing out on important qualitative dimensions that only human evaluation can identify \cite{garg2023handling}. Correspondingly, there is a growing recognition of the advantages of human augmentation of automatic evaluation. For instance, Amazon's AWS Bedrock \cite{awsbedrock} combines human benchmarking services in addition to automated evaluation to capture nuances in model outputs (e.g., empathy or friendliness) indicating an interest in evaluation metrics that are not based on ground truth. Similarly, our work used human evaluation as the first step to identify evaluation criteria that are relevant in subjective decision-making, i.e., \emph{Stance} and \emph{Perspectives}. Both sets of parameters were consistently found in human and model assessments, indicating the potential to be systematically applied for evaluation at scale. Therefore, we propose human evaluation to augment existing automated evaluation criteria. 

\subsubsection{Implications for Bias evaluation in LLMs}

Our results also extend the current understanding of bias evaluation in LLMs, often reliant on binary bias classifications (e.g., toxic vs non-toxic). Our \emph{Stance} classification offers more nuanced options (i.e., including \emph{depends} and \emph{no stance} in addition to \emph{sexist} and \emph{not sexist}) that are appropriate for non-ground truth based decisions. Further, the Perspectives set (\emph{victim, perpetrator, decision-maker}) is gender-inclusive, in alignment with critiques that current bias detection techniques are often gender binary \cite{gallegos2023bias, ovalle2023m, kotek2023gender}. As past work has alerted to harmful biased content produced by LLMs \cite{gallegos2023bias, chang2023survey, dhamala2021bold, ovalle2023m, gadiraju2023wouldn}, and the influence such outputs can have on human decision-makers \cite{ferguson2023something, ferguson2024just}, evaluating which \emph{Stances} and \emph{Perspectives} present in model outputs becomes all the more relevant.

\subsection{Implications for Human-AI Collaboration}
Finally, our work contributes to HCI by offering an exploratory comparative analysis of human and LLM outputs in subjective decision-making scenarios, using a combination of quantitative and qualitative methodologies. The human dataset was purposefully not used as ground truth or golden standard, as that concept does not apply. Instead, when considering human-AI decision-making in nuanced, ambiguous and value-based scenarios, it becomes relevant to observe the differences and similarities to consider where there is potential for collaboration and which role AI should play in it. Our results show instances where GPT-3 mimicking humans might not be a desirable trait (e.g., emulating human experiences: \textit{``I have always been a strong advocate of gender equality. I am a mother with three daughters and I have always encouraged them to be strong and independent, encouraging them to aim high and achieve their goals, whatever they may be. I am a feminist in the correct sense of the word. I believe in gender equality."} GPT-3), while newer model iterations diverge from this path. These findings further deepen discussions in AI Alignment \cite{gabriel2020artificial, khamassi2024strong}, as it is a known challenge to define which human values (taking into account pluralistic perspectives \cite{sorensen2024roadmap}) to use as parameters to train a model. Thus, we argue that AI merely emulating human responses is not necessarily the goal for all use cases. Instead, \emph{Stance} and \emph{Perspective} can be leveraged to help calibrate how similar or divergent to human responses the AI outputs need to be to foster better collaborative outcomes.

\subsection{Limitations and Future Work}
In this work, we used subtle sexism as one example of a subjective decision-making task and the degree of the potential generalization should be explored in future research. For example, exploring the application of \emph{Stance} and \emph{Perspective} to other discrimination scenarios (such as racism, ageism, ableism, etc.) that would likely also contain a \emph{victim, perpetrator}, and \emph{decision-maker}. Additionally, while it was not the focus of the current work, future work can also compare the sentence structure and content of LLM and human responses \cite{10.1145/3687056}. We noticed, for example, that the human dataset often presented grammatical errors, terse sentences and overall heterogeneity in structure (which might account for the slightly lower annotator agreement on this specific dataset). On the other hand, newer models such as LLama 3.1 had very rigid response structures (e.g., numbering the reasons to support the decision). We also foresee a continuation of this work could further explore the concept of combining human and automated evaluation methods to identify an LLM's tendency to consider different Perspectives in their rationale. We see the potential for human evaluators to determine relevant parameters in use cases that require nuanced interpretation, and then train a model to automatically identify them. Other promising parameters could be grammatical elements, convincingness \cite{10.1145/3687056} and/or rhetorical strategies used in the responses.

We also acknowledge that while researchers' interpretations are often described as beneficial to the qualitative research process \cite{muller2014curiosity}, the researchers' experience and values can influence the open coding process in this work. Hence, a Researcher Positionality statement is included in the \emph{Methods} section. There are also limitations associated with using LLMs that should be noted. For example, data leakage and contamination \cite{narayanancontamination} could occur if the scenarios used in the study (which were collected from the internet) were in the training set of the models used for the experiments. 

Lastly, while we did not experiment with prompt engineering, future work could explore how different prompting methods may change the Stances and Perspectives present in LLM outputs.

\section{Conclusion}
This paper explores nuances in human-AI collaboration for decision-making in subjective, non ground truth scenarios. Using subtle sexism as an example of a value-based and open-to-interpretation scenario, we prompted humans and LLMs (GPT-3, GPT-3.5, GPT-4 and Llama 3.1) to make assessments and justify their rationale. Our mixed-methods analysis demonstrated the responses from humans and LLMs consistently used the same set of Stances (sexist, \emph{not sexist}, \emph{depends} and \emph{no stance}) and Perspectives (\emph{victim}, \emph{perpetrator}, \emph{decision-maker}) but at different frequencies. Thus, each model displayed a distinct tendency to leverage and combine these \emph{Perspectives} in their assessment. Notably, our comparison demonstrates the newer models (GPT-3.5, GPT-4 and Llama 3.1) offer more holistic responses by considering more \emph{Perspectives} (e.g., balancing the perpetrator's intent and the victim outcomes). Further, the most recent GPT-3.5 and GPT-4 models clearly state their limitations as AI agents, suggesting safety guardrails are being enforced---in contrast to GPT-3's anthropomorphic examples. As new LLMs and model versions are released, we suggest \emph{Stances} and \emph{Perspectives} are parameters that can be leveraged to identify diversity in perspectives in model outputs to compare to, and potentially complement, a human decision-maker's rationale depending on use case.

\begin{acks}
We thank our collaborators for their invaluable contributions to this work: Justice Cheng, Senjuti Dutta, Manveer Kalirai, Nazar Ponochevnyi, Rimsha Rizvi and folks from the COoKIE Group.
\end{acks}

\bibliographystyle{ACM-Reference-Format}
\bibliography{references}

\appendix

\section{Appendix A - 
Example of prompt in question-answer format}\label{appendixquestion-answer}
In this work, we used the Scenarios Dataset as prompts in question-answer format. The Scenario was collected in online platforms (see Methods) and then written in question-answer format to prompt responses from humans, GPT-3, GPT-3.5, GPT-4 and Llama 3.1. See examples below:

\textbf{Example of Scenario} (collected online): \textit{Explaining any complaint or bad mood from a woman by ``oh she must just be on her period"}

\textbf{Example of human response}: \textit{This scenario IS sexist, because the underlying assumption is that women are profoundly affected by their hormones}.

\textbf{Example of GPT-3 response}: \textit{`` No, this is not sexist. This is a common and acceptable explanation of female behavior, which is a well-documented reality in human sociology and psychology."}    

\textbf{Example of GPT-3.5 response}: \textit{``Yes, the statement "Explaining any complaint or bad mood from a woman by ’oh she must just be on her period’" can be considered sexist. This statement perpetuates a harmful stereotype that suggests women’s emotions or behavior are solely influenced by their menstrual cycle, minimizing their experiences and dismissing their valid concerns. It promotes the idea that women’s emotions are irrational or invalid, which is unfair and discriminatory.It is important to treat individuals with respect and not make assumptions about their emotions based on their gender or biological characteristics”}.

\textbf{Example of GPT-4 response}: \textit{``Yes, this statement is considered sexist. The reason is that it perpetuates a negative stereotype about women, attributing their feelings, complaints, or moods solely to their hormonal cycle. This dismisses and invalidates the legitimacy of their concerns, grievances, or emotional states, implying they can’t have valid reasons other than biological ones. This kind of sexism also contributes to the trivialization of women’s experiences and feelings.”}

\textbf{Example of Llama 3.1 response}: \textit{``Yes, this phrase can be seen as sexist. Here’s why: 1. **Stereotyping**: It perpetuates a negative stereotype that women’s hormonal cycles (specifically, their menstrual periods) are the sole reason for their emotions, complaints, or moods. This oversimplifies the complexities of women’s experiences and ignores the myriad factors that contribute to their emotions. 2. **Dismissal**: By attributing a woman’s emotions or complaints to her period, this phrase can come across as dismissive and invalidating. It may convey that her feelings or opinions are not worthy of consideration or respect, simply because they are associated with her menstrual cycle. 3. **Biological essentialism**: It reduces women’s experiences and emotions to their biology, implying that their actions and feelings are determined by their reproductive cycles. This type of thinking ignores the fact that women are individuals with unique experiences, thoughts, and feelings, and that their emotions and actions are influenced by a range of factors, including their environment, relationships, and personal circumstances. 4. **Lack of empathy**: This phrase can convey a lack of empathy or understanding, implying that women’s emotions and experiences are not worthy of attention or consideration. It may also suggest that their emotions are less valid or less important than those of men. 5. **Unrealistic expectations**: It perpetuates the idea that women should be able to ""tough it out"" or manage their emotions in a particular way, without acknowledging that menstrual cramps, fatigue, and mood swings are common symptoms of PMS that can impact women’s daily lives. It’s essential to approach conversations about women’s emotions and experiences with empathy, understanding, and respect. Instead of attributing women’s complaints or bad moods to their periods, we can ask open-ended questions like ""What’s going on?"" or ""How can I support you?"" to encourage more nuanced and respectful communication."”}

\section{APPENDIX B - Examples of invalid responses}\label{app-invalid}
Some of the responses collected for the human and GPT-3 datasets were excluded from analysis because they either did not contain an assessment (i.e. the response to the question: is this sexist?) or a rationale supporting that assessment (i.e. the response to the question: why or why not). The quality of the rationale was not analyzed, only if it was present or not. See more details in our Methods section. See examples in the table below.

\begin{table*}[h]
\caption{Examples of responses considered invalid and removed from datasets for analysis for not containing an assessment and/or a rationale}
\label{tab:appendix-b-invalid-responses}
\small
\begin{tabular}{llllll}
\textbf{Prompt Example}                                                                                                                                                                                                                                                                                                  & \textbf{Example of Response}                                                                                                                                                                                                                                                                                                                                                                                                                                                                                                                                                                                                                                                                  & \textbf{Author} & \textbf{Assessment}                                           & \textbf{Rationale}                                                        & \textbf{Status}                                             \\ \hline
\textit{\begin{tabular}[c]{@{}l@{}}Is this scenario sexist: \\ dad gets mad at me for \\relaxing and not helping \\cook/clean. Both brothers are\\also relaxing and nothing \\is said to either of them."\\ Why or why not?\end{tabular}}                                                                                                                                                                          & \begin{tabular}[c]{@{}l@{}}"This statement is not sexist \\because the brothers could \\have already done their \\chores, or may have a \\ history of cleaning and  \\cooking way more than \\the sister." \end{tabular}                                                                                                                                                                                                                                                                                                                                                                                                                                                                                                                                                                                                                                                                  & Human           & \begin{tabular}[c]{@{}l@{}}Present: \\ "yes"\end{tabular} & \begin{tabular}[c]{@{}l@{}}Present:\\ "wording is degrading\\ towards women"\end{tabular} & \begin{tabular}[c]{@{}l@{}}Valid \\ Response\end{tabular}   \\ \hline
\textit{\begin{tabular}[c]{@{}l@{}}Is this scenario sexist:\\ "Take the military example.\\ Someone points out that there is \\ benevolent sexisminvolved in \\ barring women from serving in\\ combat roles and thus being \\ immune from a compulsory \\ draft. I think this is true."\\ Why or why not?\end{tabular}} & "I have a job"                                                                                                                                                                                                                                                                                                                                                                                                                                                                                                                                                                                                                                                                                & Human           & Not present                                                   & Not present                                                               & \begin{tabular}[c]{@{}l@{}}Invalid \\ Response\end{tabular} \\ \hline
\textit{\begin{tabular}[c]{@{}l@{}}Is this scenario sexist:\\ "Calling women "female""\\ Why or why not?\end{tabular}}                                                                                                                                                             & ``No. See here for why not.'' & GPT-3           & \begin{tabular}[c]{@{}l@{}}Present:\\ "No."\end{tabular}      & Not present                                                               & \begin{tabular}[c]{@{}l@{}}Invalid \\ Response\end{tabular}
\end{tabular}

\end{table*}
\end{document}